\documentclass{article}

\usepackage{amssymb,amsfonts,amsmath,stmaryrd}
\usepackage{cite,enumerate,float,indentfirst}
\usepackage{color}
\usepackage{comment}

\def\be{\begin{eqnarray}}
\def\ee{\end{eqnarray}}
\def\nn{\nonumber}

\definecolor{red}{rgb}{1,0,0}
\definecolor{orange}{rgb}{1,0.5,0}
\definecolor{violet}{rgb}{0.7,0,1}

\hoffset= - 1.0in         

\begin{document}

\title{\vspace{1cm}{\Large {\bf Perspectives of differential expansion}\vspace{.2cm}}
	\author{\ {\bf L. Bishler$^{a,b,c}$}, \ {\bf A. Morozov$^{b,c,d}$}
	\date{ }
}
}

\maketitle

\vspace{-5.5cm}

\begin{center}
	\hfill MITP/TH-07/20
	
	\hfill ITEP/TH-08/20
	
	\hfill IITP/TH-07/20
	
	\hfill FIAN/TH-11/20
\end{center}

\vspace{3.0cm}

\begin{center}
	
	$^a$ {\small {\it Lebedev Physics Institute, Moscow 119991, Russia}}\\
	$^b$ {\small {\it ITEP, Moscow 117218, Russia}}\\
	$^c$ {\small {\it Institute for Information Transmission Problems, Moscow 127994, Russia}}\\
	$^d$ {\small {\it Moscow Institute of Physics and Technology, Dolgoprudny 141701, Russia }}
\end{center}

\vspace{1cm}

\centerline{ABSTRACT}

\bigskip

{\footnotesize
We outline the current status of the differential expansion (DE)
of colored knot polynomials i.e. of their $Z$--$F$ decomposition into representation--
and knot--dependent parts.
Its existence is a theorem for HOMFLY-PT polynomials
in symmetric and antisymmetric representations,
but everything beyond is still hypothetical --
and quite difficult to explore and interpret.
However, DE remains one of the main
sources of knowledge and calculational means in
modern knot theory.
We concentrate on the following subjects:
applicability of DE to non-trivial knots,
its modifications for knots with non-vanishing defects
and DE for non-rectangular representations.
An essential novelty is the analysis of a more-naive ${\cal Z}$--${F_{Tw}}$ decomposition
with the twist-knot $F$-factors and non-standard ${\cal Z}$-factors
and a discovery of still another {\it triangular} and {\it universal}
transformation $V$, which converts  $\cal{Z}$ to the standard $Z$-factors $V^{-1} {\cal Z}= Z$ and allows to calculate $F$ as $F = VF_{Tw}$.

}

\bigskip

\bigskip

\section{Introduction}

Knot polynomials are gauge invariant observables in
$3d$ topological Chern-Simons (CS) theory \cite{CS},
and thus they lie on the way from the well studied conformal blocks in $2d$ \cite{CFT,AGT}
to the still-mysterious confinement-controlling Wilson loops in $4d$ QCD \cite{confin}.
This is enough to explain their central role in today's mathematical physics.
By definition the normalized HOMFLY-PT polynomial \cite{knotpols,Wit}
\be
H_R^{\cal K}(q,A) :=
\frac{1}{d_R(N)}\left< {\rm Tr}_R P\exp\left(\oint_{\cal K} {\cal A}\right)\right>
_{{\rm CS}(N,k)}
\label{aveH}
\ee
is a function of the contour (knot) ${\cal K}$, the rank of the gauge group $SU_N$,
its representation (Young diagram) $R$, of quantum dimension $d_R(N)$
and the CS coupling constant $k$.
This average is a polynomial in
peculiar variables $q = \exp\left(\frac{2\pi i}{N+k}\right)$ and $A=q^N$.
{\bf Differential expansion} is a further statement (conjecture) \cite{IMMMfe,evo,diffarth}
that there is a separation of ${\cal K}$ and $R$ variables:
\be
H_R^{\cal K}(q,A) \ \stackrel{?}{=}\ 1 + \sum_{Q \in M_R} Z_R^Q(q,A)\cdot F_Q^{\cal K}(q,A)
\label{de}
\ee
Moreover, the knot-independent $Z$-factors are actually made from products of
quantum numbers (combinatorial factors) and
``differentials'' \cite{DGR} $D_n := \{Aq^n\}=[N+n]\cdot\{q\}$, with $\{x\}:=x-x^{-1}$
and quantum numbers $[n]:=\frac{\{q^n\}}{\{q\}}$.
This statement is a representation-theory theorem for (anti)symmetric $R$ \cite{symthe} --
provided one trusts the equivalence of the definition (\ref{aveH}) and
Reshetikhin-Turaev ${\cal R}$-matrix formalism \cite{RT} in its modified form
\cite{MMMrt}, adjusted for the needs of the knot calculus for non-trivial representation $R$
(``$R$-colored''\ polynomials).
However, the simple arguments of \cite{symthe} are not sufficient for more complicated $R$,
with more than a single line or row -- and it is still disputable whether decomposition
like (\ref{de}) exists in the general case (thus the question mark over the equality).
In the present letter we briefly summarize
some of the existing evidence in favour of this conjecture.
We cover the following topics.

The first topic is the choice of the summation domain $M_R$ in (\ref{de}).
In the case of (anti)symmetric representations $R$ it consists of all the Young sub-diagrams of $R$,
what looks very nice.
The question is what happens for arbitrary $R$.
Best understood for now is the case of twist knots, where naively
$M_R^{\rm twist}=R\otimes \bar R$, i.e. is a combination of {\it composite} representations.
The strange-looking conjecture of \cite{M20} is that
supposed universality of (\ref{de}) implies that $M_R$ is just the same for all other(!) knots.
In fact, this is a little less strange than seems,
because for {\it all} rectangular  representations $R$ the product $R\otimes \bar R$
consists only of {\it diagonal} composites, and is actually equivalent for the set of
{\it sub-diagrams}.  In this letter we provide an illustration and present coefficients $F$ of a three-bridge knot.
However, this is no longer true for non-rectangular $R$, and the choice of $M_R$ in this case is still an open problem.

The second topic is a computation of coefficients $F$ from (\ref{de}) for particular knots. It is known \cite{Konodef} that even for (anti)symmetric representations coefficients $F$ obtain poles in the case of knots with non-vanishing defects. Also, we do not know HOMFLY-PT polynomials for sums of non-diagonal composite representations, which  enter the  ``moduli space'' $M_R$ in the case of non-rectangular representations $R$. It makes the search for $F$  very challenging. In this letter we apply a very powerful approach --- the use of ``families'' of knots to find $F$. Together with the $U$-matrix approach they allowed us to express coefficients $F$ via coefficients of twist knots  $F_{Tw}$ and the new universal triangular matrix $V$.

The third topic is the  shape of $Z$-factors. We seem to know $Z$-factors  for (anti)symmetric \cite{symthe}, rectangular representations \cite{KM16} and representations corresponding to the Young diagram  $[r,1]$ \cite{Morr1}. In this letter we provide some preliminary evidence on how  $Z$-factors for more general family of representations look like. We show how to investigate DE for $R = [r_1, r_2]$ and present the first members of the decomposition.

We strongly believe, that further efforts should be applied for the study of
differential expansion (\ref{de}), and we hope that, despite being tedious,
it will attract attention that it deserves.

\section{The basics of DE \label{basics}}

\subsection{Symmetric representations}

For symmetric representations $R$ the summation domain $M_{[r]}$ consists only
of symmetric representations $[i]$ with $i\leq R$,
and, remarkably, in this case the product $[r]\otimes\overline{[r]}$ consists only of
{\it diagonal} composites $([i],[i])$, thus the two realizations of the {\it set}
$M_{[r]}$ are equivalent:
\be
M_{[r]} = [r]\otimes\overline{[r]} = \oplus_{i=0}^{r} ([i],[i]) \cong \oplus_{i=0}^r [i]
\label{symmMr}
\ee
and
\be
H_{[r]}^{\cal K} = 1+\sum_{i=1}^{r} Z_{[r]}^{[i]} \cdot F_{[i]}^{\cal K}
\label{DEsym}
\ee
with the following coefficients:
\be
Z^{[i]}_{[r]}(q,A) := \prod_{j=0}^{i-1}\frac{[r-j]}{[j+1]} \{Aq^{r+j}\}\{Aq^{j-1}\}
\ \ \ \text{and} \ \ \  F_{[i]}^{\cal K}(q,A) = \frac{G_{[i]}^{\cal K}(q,A)}{\prod_{j=1}^{i-1}\{Aq^{j-1}\}}.
\label{DEsymcoef}
\ee
 For symmetric representations $R=[r]$ there is exactly one new coefficient $F_{[r]}^{\cal K}$
for every new $r$, thus for every knot they can be recursively extracted from
expression for symmetric HOMFLY-PT $H_{[r]}^{\cal K}$ polynomials.

DE (\ref{DEsym}) is a direct corollary of representation theory, namely
the equivalence of antisymmetric representations:
\be
H_{[1^k]}^{\cal K}(q,A=q^N)=H_{[1^{N-k}]}^{\cal K}(q,A=q^N),
\ee
i.e. for the group $SU_N$.
Relation to symmetric representations is provided by the symmetry
\be
H_R (A, q^{-1}) = H_{R^\vee}(A, q),
\label{transp}
\ee
where $R^\vee$ denotes transposed Young diagram $R$.
Simultaneous inversion of both $A$ and $q$ changes the knot to its mirror.
To (\ref{transp}) one should add triviality of all symmetric characters
for the case of $SU(1)$, i.e. for $A=q$:
\be
 H_{[r]}^{\cal K}(q,A=q) = 1
\ \ \Longrightarrow \ \ H_{[r]}^{\cal K} -1 \sim \{A/q\} \ \
\Longrightarrow H_{[1^r]}^{\cal K} - 1 \sim \{Aq\}.
\ee
Since we deal with the {\it normalized} HOMFLY-PT, antisymmetric ones
do not need to be unities at $A=q$,
because they include division by dimension, which vanishes for $A=q$ and $R\neq [r]$.

\bigskip

{\footnotesize
To make the story complete, we repeat and extend here the argument of \cite{symthe}.
\begin{itemize}

\item{
For $R=[1]$ we have $H_{[1]}^{\cal K}(q,A=q) =1 \ \ \Longrightarrow \ \ \
H_{[1]}^{\cal K}(q,A) - 1\  \vdots\  \{A/q\}$.
Since in this case the diagram is symmetric, eq.(\ref{transp}) implies invariance
under the $q$ inversion, and in fact
\be
H_{[1]}^{\cal K}(q,A) = 1 + \{Aq\}\{A/q\} \cdot F_{[1]}^{\cal K}(q,A)
\ee
with some Laurent polynomial $ F_{[1]}^{\cal K}(q,A) = F_{[1]}^{\cal K}(q^{-1},A)$.
}

\item{
The idea of differential expansion \cite{IMMMfe} comes from the basic property
of ``special'' polynomials, which are HOMFLY-PT, evaluated at at $q=1$:
\be
{\cal H}_R^{\cal K}(q=1,A) = \left({\cal H}_{[1]}^{\cal K}(q=1,A)\right)^{|R|}
= \left(1+\{A\}^2\cdot F_{[1]}^{\cal K}(q=1,A)\right)^{|R|}
= \sum_{I=0}^{|R|} \frac{|R|!}{I!(|R|-I)!}\cdot F_{\cal K}^I\cdot\{A\}^{2I}
\label{speDE}
\ee
which is just a binomial expansion of degree $|R|$ in even powers of $\{A\}$.
DE is the question, of what is the ``quantization'' of this property for $q\neq 1$.
For brevity we substituted $F_{\cal K}(A) := F_{[1]}^{\cal K}(q=1,A)$
}

\item{
Next, for $R=[2]$ and $R=[2]^\vee = [1,1]$ we have two interesting relations, for
$N=2$ and $N=3$:
\be
H_{[1,1]}^{\cal K}(q,A=q^2) = 1 \ \ \Longrightarrow \ \ \
H_{[1,1]}^{\cal K}(q,A) - 1 \  \vdots\ \{Aq\} \{A/q^2\}
\nn
\ee
\vspace{-0.6cm}
\be
\!\!\!\!\!\!\! \!\!\!\!\!\!\!
H_{[1,1]}^{\cal K}(q,A=q^3) = H_{[1]}^{\cal K}(q,A=q^3)
= 1 + F_{[1]}^{\cal K}(q,A=q^3)\{q^4\}\{q^2\}
\ \ \Longrightarrow \ \ \
H_{[1,1]}^{\cal K}(q,A) - 1 - F_{[1]}^{\cal K}(q,A=q^3)\{q^4\}\{q^2\} \  \vdots\  \{A/q^3\}
\nn
\ee
Together they imply that
\be
H_{[1,1]}^{\cal K}(q,A)
= 1 + \tilde {G}_{[1,1]}^{\cal K}(q,A) \{Aq\}\{A/q^2\}
= 1 + [2]F_{[1]}^{\cal K}(q,A )\{Aq\}\{A/q^2\} + G_{[1,1]}^{\cal K}(q,A)\{Aq\}\{A/q^2\}\{A/q^3\}
\ee
for some Laurent polynomial $G_{[1,1]}^{\cal K}(q,A)$.
By transposition symmetry
\be
H_{[2]}^{\cal K}(q,A)
= 1 + [2]F_{[1]}^{\cal K}(q,A )\{Aq^2\}\{A/q\} + G_{[2]}^{\cal K}(q,A)\{Aq^3\}\{Aq^2\}\{A/q\}
\ee
with $G_{[2]}^{\cal K}(q,A) = G_{[1,1]}^{\cal K}(q^{-1},A)$

}

\item{
In the same way, one can get:
\be
\!\!\!\!\!
H_{[r]}^{\cal K} - H_{[r']}^{\cal K} \sim \{Aq^{r+r'}\}\{A/q\} \ \ \ \Longrightarrow \ \ \
H_{[r]}^{\cal K}(q,A)
= 1 + [r]F_{[1]}^{\cal K}(q,A )\{Aq^r\}\{A/q\} + \sum_{i=2}^r G_{[i]}^{\cal K}(q,A)\cdot \{A/q\}
\cdot \prod_{j=0}^{i-1} \frac{[r-j]}{[j+1]} \{Aq^{r+j}\}
\ee
for the set of Laurent polynomials $G_{[i]}^{\cal K}(q,A) = G_{[1^i]}^{\cal K}(q^{-1},A)$.
To prove this one should use identities like $[r]D_r-[r']D_{r'} = [r-r']D_{r+r'}$.
Note that $G^{\cal K}_{[i]}(q=1,A)$ should vanish with $i\geq 2$, according to (\ref{speDE}),
because the power of differentials is $i+1 < 2i$, actually, it follows that
$G^{\cal K}_{[i]}(q,A=q^N)\sim \{q\}^{i-1}$ as $q$ approaches $1$.

}

\end{itemize}

}

\subsection{Defect of the differential expansion \cite{Konodef}}

According to the above definition (\ref{DEsymcoef}), the coefficients $F_{[i]}$ of the differential expansion
are not Laurent polynomials --- only $G_{[i]}$ need to be such, which was shown with the use of representation-theory arguments.
However, there is a set of knots for which $F_{[i]}$ are Laurent polynomials.
This set is characterized by vanishing of the peculiar quantum number --- {\it defect}:
$\ \delta^{\cal K}=0$.
Such polynomials have $q$-independent coefficient $F_{[1]}(A)$,
and therefore their fundamental Alexander polynomials are just
\be
{\rm Al}_{[1]}^{{\cal K}^{(0)}} = 1 + {\rm const}^{\cal K}\cdot \{q\}^2
\ee
i.e. Laurent polynomials of degree $1$ in $q^{\pm 2}$
(the constant coefficient depends on ${\cal K}$).
It turns out that for any concrete knot ${\cal K}$ the power of
its Alexander polynomial ${\rm deg}\left({\rm Al}_{[1]}^{\cal K}\right) = 1+\delta^{\cal K}$
defines the divisibility of $G_{[i]}$ by a somewhat mysterious rule \cite{Konodef}:
\be
G_{[i]}^{\cal K}(q,A) = F_{[i]}^{\cal K}(q,A)\ \cdot \!\!\!\!\!\!\!\!
\prod_{j=1}^{{\rm entier}\left(\frac{i-1}{1+\delta^{\cal K}}\right)}
\!\!\!\!\{Aq^{j-1}\}
\ee
For defect $\delta^{\cal K}=-1$ Alexander polynomial is unity,
i.e. $F_{[1]}^{{\cal K}^{-1}}\!\!\!=0$,
but all other $F_{[i]}$ are just polynomials, as for defect $0$.

Defect-zero knots are relatively rare,
the most important series consist of twist and
double-braid knots, the first example beyond these series
is pretzel $9_{46}$.

\subsection{Rectangular  representations}

Clearly, the knowledge of symmetric $H_{[r]}^{\cal K}$ for a given knot
is sufficient to get all the $F_{[r]}^{\cal K}$,
by recursively using (\ref{DEsym}) with known $Z_{[i]}^{[r]}$.
The case of rectangular representations $R=[r^s]$ is more difficult.
The summation domain $M_{[r^s]}$ consists of {\it all} Young sub-diagrams $\lambda \in [r^s]$,
which do not need to be rectangular:
\be
M_{[r^s]} =[r^s]\otimes \overline{[r^s]} =  \sum_{ \lambda \in [r^s] } (\lambda, \lambda)
\ \ \ \ \ \ \ {\rm and} \ \ \ \ \ \ \
H_{[r^s]} = \sum_{\lambda \in [r^s] } Z_{[r^s]}^{\lambda}\cdot F_{\lambda}.
\label{rec}
\ee
Also the representation theory arguments are no longer sufficient to {\it justify} the DE,
still the argument with the {\it special} polynomials persists.

\bigskip

{\footnotesize
This time we have the relations
\vspace{-0.3cm}
\be
H^{\cal K}_{[r^s]} - H^{\cal K}_{[r^{s'}]} \ \vdots \ \{A/q^{s+s'}\}
\ \ \ \ \ {\rm and} \ \ \ \ \ \ \
H^{\cal K}_{[{r}^s]} - H^{\cal K}_{[{r'}^s]} \ \vdots \ \{Aq^{r+r'}\}.
\ee
For example, this means that
\vspace{-0.3cm}
\be
H_{[2,2]}^{\cal K} -1 \ \vdots \ \{Aq^2\}\{A/q^2\}, \ \ \ \ \ \ \
H_{[2,2]}^{\cal K} - H_{[2]}^{\cal K}\ \vdots \ \{A/q^3\} \ \ \ \ \ {\rm and} \ \ \ \ \
H_{[2,2]}^{\cal K} - H_{[1,1]}^{\cal K} \vdots \ \{Aq^3\}.
\ee
Substituting $H_{[2]}^{\cal K} = 1+[2]F_{[1]}^{\cal K}\{Aq^2\}\{A/q\}
+ F_{[2]}^{\cal K}\{Aq^3\}\{Aq^2\}\{A\}\{A/q\}$ we get
\be
H^{\cal K}_{[2,2]} = 1 + [2]^2F_{[1]}^{\cal K}\{Aq^2\}\{A/q^2\}   +
[3] F_{[2]}^{\cal K} \{Aq^{3}\}\{Aq^2\}\{A/q\}\{A/q^2\}   +
[3] F_{[1,1]}^{\cal K} \{Aq^2\}\{Aq\}\{A/q^2\}\{A/q^3\}    +
\nn
\ee
\vspace{-0.5cm}
\be
+\tilde F_{[2,2]}^{\cal K}\{Aq^3\}\{Aq^2\}\{A/q^2\}\{A/q^3\},
\label{DEder21}
\ee
since  $[2]\{A/q^2\}-\{A/q\} = \{A/q^3\}$  and $[3]\{A/q^2\}-\{A\} = [2]\{A/q^3\}$.
However, this is not enough to justify the further separation
\be
\tilde F^{\cal K}_{[2,2]} = [2]^2F^{\cal K}_{[2,1]} \cdot\{A q\}\{A/q\} + F^{\cal K}_{[2,2]} \cdot \{A q^2\} \{A q\} \{A/q\} \{A/q^2\}
\ee
in the second line. 

A direct generalization of (\ref{DEder21}) is
\be
H^{\cal K}_{[r^s]} = 1 + [r][s]F_{[1]}^{\cal K}\{Aq^r\}\{A/q^s\}
+ \frac{[r][r-1][s][s+1]}{[2]^2} F_{[2]}^{\cal K}\{Aq^{r+1}\}\{Aq^r\}\{A/q^{s-1}\}\{A/q^s\}
  \  +
\nn
\ee
\vspace{-0.5cm}
\be
+ \frac{[r][r+1][s][s-1]}{[2]^2} F_{[1,1]}^{\cal K}\{Aq^{r}\}\{Aq^{r-1}\}\{A/q^s\}\{A/q^{s+1}\}
+\ O\Big(\{Aq^{r+1}\}\{Aq^r\}\{A/q^s\}\{A/q^{s+1}\}\Big)
\label{DEderrect}
\ee
and for $r,s>2$ we actually have more conditions to further structure the remaining part in the second line,
still this reasoning by itself is insufficient to justify the full-fledged DE.

}

\bigskip

 $Z$-factors for all representations $\lambda \in [r^s]$
have a very simple form \cite{KM16}:
\be
Z^\lambda_{[r^s]} = d_{\lambda^\vee}(r) d_\lambda(s)
\prod_{\Box\in \lambda} \Big\{Aq^{a'_\Box-l'_\Box+r}\Big\}
  \Big\{Aq^{a'_\Box-l'_\Box-s}\Big\}
  \label{recZ}
\ee
where $d_{Q}(m)$ is a quantum dimension of representation $Q$ at the point $A=q^m$, i.e. the Schur polynomials at the topological locus $\chi_{Q}\{p_k^* =[m k]/[k]\} $.
As in the symmetric case, the coefficients $F_\lambda^{\cal K}$ are expected to be
Laurent polynomials in $q$ and $A$ only for defects $\delta^{\cal K}=0$ and $-1$.
These coefficients are explicitly found for all twist and double-braid knots in \cite{M16}
and \cite{KNTZ},
but the statement remains true beyond these families.
For example, for a three-bridge knot $9_{46}$:
{\footnotesize
\be
F_{[1]}^{[9_{46}]}=A^{-4}\cdot\big(1 + A^2\big) , \ \ \ \ \ \ \ \ \ \
F_{[2]}^{[9_{46}]}= q^{-8}A^{-8}\cdot \Big(1 + A^2q^2 + A^2q^4 + A^4q^8\Big), \nn \\
F_{[3]}^{[9_{46}]}=q^{-24}A^{-12}\Big(1 + A^2q^4\Big)\Big(1 + A^2q^6 + A^2q^8 - A^4q^{10}
+ A^4q^{14} + A^4q^{16}\Big). \ \ \
\nn
\ee
}

\noindent
See \cite{sleptsov} and \cite{Kostud} for generalization of these  formulas to all symmetric representations.
The fact that they exist for non-symmetric representations too is more important for our purposes, e.g.
\be
F_{[2,1]}^{[9_{46}]}=
\frac{1}{A^{12}}\left\{\frac{[10]}{[5][2]}\Big(A^6+[3]A^4\Big)+\frac{[6]}{[2]}A^2+1+[3]^2\{q\}^4 A^4\right\},
\nn \\ \nn \\
F_{[2,2]}^{[9_{46}]}= \frac{1}{A^{16}}\left\{ \frac{[20][2]}{[10][4]}A^8 + \frac{[10][2]}{[5]}A^6
+ \frac{[4][3]}{[2]}A^4+[2]^2A^2+1\right\}.
\label{F94621}
\ee

\subsection{Non-rectangular representations}

Since we know $F_{[2,1]}$, we can also hope to describe non-rectangular $R=[2,1]$.
However, in this case the {\it multiplicities} occur, the $Z$-factors are no longer
given by (\ref{recZ})  and the space $M_R$ consists
not only of sub-diagrams of $R$.
In the simplest example of $R=[2,1]$ the  $Z$-factors are
\be
Z_{[2,1]}^{[1]} = D_1D_{-1}+(q^2+q^{-2})D_2D_{-2}, \ \ \ \ \ Z_{[2,1]}^{[2]}=\frac{[3]}{[2]}D_3D_2D_0D_{-2},
\ \ \ \ \  Z_{[2,1]}^{[2,1]}= D_{3}D_2D_1D_{-1}D_{-2}D_{-3}
\label{Z21}
\ee
and there is one new $Z_{[2,1]}^{X2}=-[3]^2\{q\}^4D_2D_{-2}$ with associated $F$-function
 \be
F_{X2}^{9_{46}} =
\frac{1}{A^8}\left\{\frac{[6]^2}{[2]^2}-\frac{[14]}{[7][2]}A^2-\{q\}^2A^4\right\}.
\label{F946X2}
\ee
The ambiguity in beyond-symmetric DE manifests itself in the possibility of simultaneous shift
\be
\delta F_{[2,2]} = -\rho \cdot [2]^2[3]^2\{q\}^4,  \ \ \ \
\delta F_{[2,1]} = \rho\cdot [3]^2\{q\}^4 D_2D_{-2}, \ \ \ \
\delta F_{X2} = \rho \cdot D_3D_2D_1D_{-1}D_{-2}D_{-3}
\ee
with arbitrary common Laurent polynomial $\rho(q,A)$.
It actually involves also the shifts of {\it higher} $F_X$ with $X>[2,2]$,
which can easily get singular if $\rho$ is chosen inappropriately.
We, however, did not look at these restrictions and therefore our particular guess (\ref{F94621}, \ref{F946X2}) can
turn out to be wrong when we learn more: our choice is just to ``minimize'' the expressions for $F$.
Truly important is the fact that at least some polynomial choice for $F^{9_{46}}$ does exist.

\bigskip

\noindent
{\bf Polynomial} formulas for $F^{9_{46}}$ provide {\bf the first new evidence in this letter}:
that {\bf DE remains true beyond symmetric representations --- and not only for twist and double
braid families}.

\section{Other defects: diminished $Z$-factors or non-polynomial $F$ \label{nonzerodefect}}

We now attempt to move beyond defect zero,
as we already know this implies the diminishing of $Z$-factors,
or, as we prefer to formulate it, an appearance of certain poles in the $F$-factors.
The problem is that beyond symmetric representations we do not know when
these poles appear and what they are.
In this letter we consider only defect $\delta_{\mathcal{K}} = 1$ and only representations $R=[2,1]$
and $R=[2,2]$, but even in this case the problem turns out to be rather complicated. Consideration of DE for mutant knots with defect two you can find in \cite{mutJETP}.

\subsection{Family$_a$}

A powerful method \cite{DMMSS,evo} to deal with knot polynomials is to consider particular knots  as
members of an ``evolution family'', depending on additional parameters.
For an illustration, we take a simple 2-bridge family
(``${\rm family}_a$'') of defect-one knots,
described by the arborescent formula
\be
H_R^{{\rm family}_a(k)} = d_R
\left<\bar S \bar T^2 \bar S \bar T^2 \bar S \bar T^2 \bar S \bar T^{2k} \bar S
\right>_{\emptyset\emptyset}
\label{familya}
\ee
It includes the following members of the Rolfsen table:
\be
\begin{array}{c||c|c|c|c|c|c|c|c|c}
k & \ldots & -3 & -2 & -1 & 0 & 1 & 2 & 3 & \ldots\\
\hline
{\rm family}_a(k)& & 10_4 & 8_4 & 6_2 & 3_1 & 5_1 & 7_3 & 9_4 &
\end{array}
\ee
Advantage of the evolution is that we know how the answer depends on $k$
through explicitly known eigenvalues $\lambda_Q$
\be
H_R^{{\rm family}_a(k)}(q,A) = \sum_{Q\in R\otimes\bar R}
\alpha_Q^{{\rm family}_a(k)}(q,A)\cdot \lambda_Q^{2k}
\ee
and coefficients $\alpha_Q^{{\rm family}_a(k)}(q,A)$ can be found from a few $H_R$ for ``small'' knots
in the family.
Moreover, this can be done directly for the DE coefficients $F_Q^{{\rm family}_a(k)}$.
The only problem is that the number of needed ``small'' knots increases with $R$.
Just two eigenvalues contribute into $F_{[1]}$,
$\lambda_\emptyset =1$ and $\lambda_{[1]} = -A$, thus the knowledge of just the two polynomials,
say $H_{[1]}^{3_1}$ and $H_{[1]}^{5_1}$, is enough --- and one gets
\be
F_{[1]}^{{\rm family}_a(k)} = -A^2\left(1 \ + \
(q^2+q^{-2})\cdot A^2\cdot\frac{ 1-A^{2k} }{1-A^2}\right).
\ee
For $F_{[2]}$ we need three polynomials, and get
{\footnotesize
\be
\!\!\!\!\!\!\!
F_{[2]}^{{\rm family}_a(k)} =  q^2A^4 + \frac{(1+q^{-2})A^6\Big(A^2(1+q^6)-(q^2+q^6)\Big)}{A^2-1}
\cdot\frac{A^{2k}-1}{A^2-1}
+ \frac{(q^{12}+q^8+q^6+1)A^8}{q^4A^2-1}\cdot\left(\frac{q^{4k}A^{4k}-1}{q^2A^2-1}
- (1+q^{-2})\cdot\frac{A^{2k}-1}{A^2-1}\right).
\nn
\ee
}

\noindent
Singularity at $q^4A^2-1$ in the last term is actually canceled by the bracket in the numerator,
because at $A=q^{-2}$ we have $q^4A^4 = A^2$.
However the middle term has one uncanceled  $D_0$ in denominator ---
because the defect of the knots is non-vanishing.

In full accordance with (\ref{DEder21}) we further observe that the difference
\be
{\footnotesize
H^{{\rm family}_a(k)}_{[2,2]} - \left( 1 + [2]^2F_{[1]}^{{\rm family}_a(k)} D_2D_{-2} +
[3]F_{[2]}^{{\rm family}_a(k)} D_3D_2D_{-1}D_{-2} +
[3]F_{[1,1]}^{{\rm family}_a(k)} D_2D_1D_{-2}D_{-3}\right)
}
\ee
is divisible by $D_3D_2D_{-2}D_{-3}$.
Note that the difference and the ratio are Laurent polynomials ---
despite $F_{[2]}$ and $F_{[1,1]}$ are not:
they contain a factor $D_0^{-1}$ because defect is greater than zero.
Still the two poles cancel for a very general reason.

\bigskip

{\footnotesize

\bigskip

When defect exceeds zero, $F_{[2]}=\frac{G_{[2]}}{D_0}$
and $F_{[1,1]}= \frac{G_{[1,1]}}{D_0}$ are no longer polynomials,
they acquire $D_0$ in denominator.
Still the contribution (\ref{DEderrect}) to $H_{[r^s]}$ in rectangular representations $[r^s]$,
\be
\frac{[r][r-1][s][s+1]}{[2]^2} \frac{D_rD_{r+1}D_{-s}D_{1-s}}{D_0} G_{[2]}
+ \frac{[r][r+1][s][s-1]}{[2]^2} \frac{D_rD_{r-1}D_{-s}D_{-s-1}}{D_0} G_{[1,1]}
\label{polecance211}
\ee
is always a polynomial(!).
Residue at the poles at $A=\pm 1$ are actually independent of $r$ and $s$
and equal to
\be
\left.(G_{[2]} + G_{[1,1]})\right|_{A=\pm 1} = 0
\ee
This condition means that $G_{[2]}$ changes sign under the inversion of $q\longrightarrow q^{-1}$ at the points $A=\pm 1$, while $G_{[1,1]}$
is obtained from $G_{[2]}$ by inversion of $q$.
One more way to formulate this is in terms of DE of Alexander polynomial
in representation $R=[2]$.

Further, the contribution of the size-three diagrams to the same $H_{[r^s]}$ is
\be
\frac{[r][r-1][r-2][s][s+1][s+2]}{[2]^2[3]^2}
\frac{D_rD_{r+1}D_{r+2}D_{-s}D_{1-s}D_{2-s}}{D_1D_0}\cdot G_{[3]}
+ \nn \\
+\ \frac{[r+1][r][r-1][s+1][s][s-1]}{[3]^2}
\frac{D_{r+1}D_rD_{r-1}D_{-s+1}D_{-s}D_{-s-1} }{D_1D_{-1}}\cdot G_{[2,1]}
+ \nn \\
+\ \frac{[r][r+1][r+2][s][s-1][s-2]}{[2]^2[3]^2}
\frac{D_rD_{r-1}D_{r-2}D_{-s}D_{-s-1}D_{-s-2 }}{D_0D_{-1}}\cdot G_{[1,1,1]}.
\label{dia3cont}
\ee
Residues at $A=\pm 1$ are vanishing, because of the property:
\be
\left.(G_{[3]} - G_{[1,1,1]})\right|_{A=\pm 1} = 0.
\ee
Residues at $A=\pm q^{-1}$ vanish because of relation between
$G_{[2,1]}$ and $G_{[3]}$:
\be
\left.([2] G_{[2,1]} + G_{[3]})\right|_{A=\pm q^{-1}} = 0.
\ee
Residue at $A=\pm q$ then vanishes automatically because
$G_{[1,1,1]}(A,q) = G_{[3]}(A,q^{-1})$ while $G_{[2,1]}(A,q^{-1})=G_{[2,1]}(A,q)$.

}

\bigskip

Coming back to $\tilde G_{[2,2]}^{{\rm family}_a(k)}$,
it now depends on six eigenvalues,
but coefficients should still be split between the two $F$-functions,
$F_{[2,1]}\oplus F_{[2,2]}$.
Eigenvalue $A^{8k}$ contributes only to $F_{[2,2]}$, but this fact is not sufficient to
decide how to split everything else.
Moreover the sum is proportional to $D_3D_2D_{-2}D_{-3}$ only,
but we do not know if the $D_1D_{-1}$ provides poles in both $F_{[2,1]}$
and $F_{[2,2]}$
and/or, if there can be an  extra pole $D_2D_{-2}$ in $F_{[2,2]}$.
Surprisingly or not, this problem is hard enough to be solved by the guess-and-check method
and requires a more systematic approach.
Very recently such a tool was found, and we deliberately selected our ${\rm family}_a$
to make it applicable.
Example of a single family allows to fix the ambiguities,
and we will get a conjecture, applicable to all defect-one knots
(the same strategy will then work for other representations and defects).

\subsection{The $U$-matrix approach}

The interesting option is to treat (\ref{familya}) by the same pentad method \cite{M20},
which was successfully applied to twist knots.
For this purpose we insert the auxiliary $U$-matrix,
which is already explicitly known for many $R$:\footnote{
We remind that $U$-matrix is constructed by the following chain of steps \cite{M20},
which involve only the information about {\it twist} knots.
One begins with explicitly known ${\cal B}$, expressed through skew Schur functions
(and through skew Macdonald polynomials, if one deals with hyperpolynomials).
Then one constructs its eigenfunction matrix ${\cal E}$ and properly normalize it.
From ${\cal E}$ one can build the Racah matrix $\bar S$, it is bilinear in ${\cal E}$.
Finally $U={\cal E}\bar S^{-1}$.
Matrices ${\cal B}$ and ${\cal E}$ are lower-triangular
and {\it universal}, i.e. independent of representation $R$.
$\bar S$ and $U$ do not possess these properties and need to be calculated for every $R$.
However, $\bar S$ is unitary and symmetric, and the second inclusive Racah matrix $S$
(unitary, but no longer symmetric)
is its diagonalizing matrix, i.e. is a solution of still another linear algebra problem.
Thus the entire pentad $\{{\cal B},{\cal E},U,\bar S,S\}$ can be obtained by linear algebra
operations from Schur functions, at least for all rectangular $R$.
For non-rectangular $R$ the starting formula for ${\cal B}$ is not fully available yet. Such calculations for $R = [2,1]$ are made in the section \ref{Uapp21}.
}
\be
H_R^{{\rm family}_a(k)} =
d_R\left<\bar S \bar T^2 \bar S \bar T^2 \bar S \bar T^2  \bar S \bar T^{2k+2} \bar S
\right>_{\emptyset\emptyset}
= \sum_X \underbrace{ d_R \left<\emptyset\Big|\bar S \bar T^2 \bar S \bar T^2 \bar S \bar T^2
\bar S \bar T^{-2}\bar S U\Big|X\right>}_{{\cal Z}_R^{X\,{\rm family}_a}}
\underbrace{\left<X\Big|U^{-1} \bar S \bar T^{2k+2} \bar S\Big|\emptyset
\right>}_{F_X^{{\rm twist}(k)}}
\label{familya1}
\ee
Now dependence on the evolution parameter $k$ is fully contained in
the coefficients $F_X^{{\rm twist}(k)}$, which are literally the same as for
the twist knots, but the factors ${\cal Z}_R^{X\,{\rm family}_a}$ are new and different from the standard $Z$-factors
\be
Z_R^X = d_R \left<\emptyset\Big|\bar S \bar T^2  \bar S \bar T^{-2}\bar S U\Big|X\right>
\ee
of the differential expansion.
For example,
\be
{\cal Z}_{[1]}^{\emptyset\,{\rm family}_a} = \!\!\overbrace{1}^{Z^\emptyset_{[1]}} \!\!-\
A^2\overbrace{D_1D_{-1}}^{Z^{[1]}_{[1]}},
\ \ \ \ \ \ \ \
{\cal Z}_{[1]}^{[1]\,{\rm family}_a} =   \frac{[4]}{[2]}A^2\overbrace{D_1D_{-1}}^{Z^{[1]}_{[1]}},
\ee
{\footnotesize
\be
{\cal Z}_{[2]}^{\emptyset\,{\rm family}_a} = \!\!\overbrace{1}^{Z^\emptyset_{[2]}} \!\!
- \ A^2 \overbrace{[2]D_2D_{-1}}^{Z^{[1]}_{[2]}}   \
+\ q^2A^4\overbrace{D_3D_2D_0D_{-1}}^{Z^{[2]}_{[2]}},
\ \ \ \ \ \ \ \ \ \ \ \ \ \ \ \ \ \ \ \ \ \ \ \ \ \ \ \ \ \ \ \
\ \ \ \ \ \ \ \ \ \ \ \ \ \ \ \ \ \ \ \ \ \ \ \ \ \ \ \ \ \ \ \
\ \ \ \ \ \ \ \ \ \ \ \ \ \ \ \ \ \ \ \ \ \ \ \ \ \ \ \ \ \ \ \
\nn\\
{\cal Z}_{[2]}^{[1]\,{\rm family}_a} =
 \frac{[4]}{[2]}A^2\overbrace{[2]D_2D_{-1}}^{Z^{[1]}_{[2]}}
-- \frac{[2]A^3}{D_0}\left( \frac{[4]}{[2]}q^3-\frac{[6]}{[3]}A^2q^2 \right)\overbrace{D_3D_2D_0D_{-1}}^{Z^{[2]}_{[2]}},
\ \ \ \ \ \ \ \ \ \ \ \ \ \ \ \ \ \ \ \
\nn \\
{\cal Z}_{[2]}^{[2]\,{\rm family}_a} =  A^4 q^2 \left(\frac{[9]}{[3]}+q^2\right)\overbrace{D_3D_2D_0D_{-1}}^{Z^{[2]}_{[2]}}
\nn
\ee
}

\noindent
and so on.
For the $R=[2,2]$ case

{\footnotesize
\be
{\cal Z}^{\emptyset\,{\rm family}_a}_{[2,2]} = 1 - A^2Z_{[2,2]}^{[1]}+q^2A^4Z_{[2,2]}^{[2]}
+\frac{A^4}{q^2}Z_{[2,2]}^{[1,1]}-A^6Z_{[2,2]}^{[2,1]}+A^8Z_{[2,2]}^{[2,2]},
\nn\\ \nn \\
{\cal Z}^{[1]\,{\rm family}_a}_{[2,2]}  =  \frac{[4]}{[2]}A^2 Z_{[2,2]}^{[1]}
- \frac{[2]A^3}{D_0}\left( \frac{[4]}{[2]}q^3-\frac{[6]}{[3]}A^2q^2 \right) Z_{[2,2]}^{[2]}
- \frac{[2]A^3}{D_0}\left( \frac{[4]}{[2]}\frac{1}{q^3}-\frac{[6]}{[3]}\frac{A^2}{q^2} \right)Z_{[2,2]}^{[1,1]}
+\nn \\
+ \frac{[3]A^4}{D_1D_{-1}}\Big( \frac{[8]}{[4]}A^4 -\frac{[6]^2}{[3]^2}A^2+2\frac{[4]}{[2]}-2  \Big) Z_{[2,2]}^{[2,1]}
-\frac{[8][2]^2}{[4]}\cdot A^8Z_{[2,2]}^{[2,2]},
\nn \\ \nn \\
\!\!\!\!\!\!\!\!\!\!\!\!
{\cal Z}^{[2]\,{\rm family}_a}_{[2,2]}  = A^4 q^2 \left(\frac{[9]}{[3]}+q^2\right) Z_{[2,2]}^{[2]} +
\left( 	\frac{[3]A^6}{[2]q^7}\left(\frac{\{q\}[8][6]}{[4][3]}q^{10}A^2-q^{14}-2q^8-1\right) -A^9q^7\frac{\{q\}[8][3]}{[4][2]D_1}  \right) Z_{[2,2]}^{[2,1]}
+ [3]\left(\frac{[12]}{[4]} +q^2\right)\cdot A^8  Z_{[2,2]}^{[2,2]},
\nn \\ \nn \\
\!\!\!\!\!\!\!\!\!\!\!
{\cal Z}^{[1,1]\,{\rm family}_a}_{[2,2]}  =
\frac{A^4}{q^2}  \left(\frac{[9]}{[3]}+\frac{1}{q^2}\right) Z_{[2,2]}^{[1,1]} +
\left(	-\frac{[3]q^7A^6}{[2]}\left(\frac{\{q\}[8][6]}{[4][3]}\frac{A^2}{q^{10}}+\frac{1}{q^{14}}+\frac{2}{q^8}+1\right) +\frac{A^9}{q^7}\frac{\{q\}[8][3]}{[4][2]D_{-1}}\right)
Z_{[2,2]}^{[2,1]}
+ [3]\left(\frac{[12]}{[4]} +\frac{1}{q^2}\right)\cdot A^8   Z_{[2,2]}^{[2,2]},
\nn \\ \nn \\
\!\!\!\!\!\!\!\!\!\!\!\!
{\cal Z}^{[2,1]\,{\rm family}_a}_{[2,2]}
=\left(A^2[3]\{q\}^2+\frac{[10][8][6] }{[5][4][3]}\right)\cdot A^6Z_{[2,2]}^{[2,1]}
-\frac{[10][8][6][2]^2}{[5][4][3]}\cdot A^8 Z_{[2,2]}^{[2,2]},
\nn \\
{\cal Z}^{[2,2]\,{\rm family}_a}_{[2,2]} =
A^8\cdot \left( [11]+\frac{[20]}{[4]}\right)
Z_{[2,2]}^{[2,2]}.
\nn
\ee
}

\noindent
Formulas above demonstrate that for our ``family$_a$''
we get an  upper-triangular transformation matrix $V^{{\rm family}_a}$,
which is {\it universal}  in the sense that its entries
are independent of representation $R$:
\be
{\cal Z}^{X\,{\rm family}_a}_{R} = \sum_{Y \subset R } V^{X\ {\rm family}_a}_{Y}\cdot Z^Y_{R}.
\ee
The segment of $V$, relevant for $R=[2,2]$, is

\bigskip

{\tiny
	\be
		\!\!\!\!\!\!\!\!\!\!\!\!
	V=\left(\begin{array}{cccccc}
	1 & -A^2 & q^2A^4 & \frac{A^2}{q^2} & -A^6 & A^8
	\\ \\
	0 &  \frac{[4]}{[2]}A^2 &\frac{[2]A^3}{D_0}\left( \frac{[4]}{[2]}q^3-\frac{[6]}{[3]}A^2q^2 \right)&\frac{[2]A^3}{D_0}\left( \frac{[4]}{[2]}\frac{1}{q^3}-\frac{[6]}{[3]}\frac{A^2}{q^2} \right)
	&
	\frac{[3]A^4}{D_1D_{-1}}\Big( \frac{[8]}{[4]}A^4 -\frac{[6]^2}{[3]^2}A^2+2\frac{[4]}{[2]}-2  \Big) &
	-\frac{[8][2]^2}{[4]}\cdot A^8
	\\ \\
	0&0& A^4 q^2 \left(\frac{[9]}{[3]}+q^2\right) &0&
	\frac{[3]A^6}{[2]q^7}\Big(\frac{\{q\}[8][6]}{[4][3]}q^{10}A^2-q^{14}-2q^8-1\Big) -A^9q^7\frac{\{q\}[8][3]}{[4][2]D_1}
	& [3]\Big(\frac{[12]}{[4]} +q^2\Big)\cdot A^8
	\\ \\
	0&0&0&\frac{A^4}{q^2} \left(\frac{[9]}{[3]}+\frac{1}{q^2}\right)&
	
	-\frac{[3]q^7A^6}{[2]}\left(\frac{\{q\}[8][6]}{[4][3]}\frac{A^2}{q^{10}}+\frac{1}{q^{14}}+\frac{2}{q^8}+1\right) +\frac{A^9}{q^7}\frac{\{q\}[8][3]}{[4][2]D_{-1}}
	& [3]\Big(\frac{[12]}{[4]} +\frac{1}{q^2}\Big)\cdot A^8
	\\ \\
	0&0&0&0& \left(A^2[3]\{q\}^2+\frac{[10][8][6]}{[5][4][3]}\right) A^6 &
	-\frac{[10][8][6][2]^2}{[5][4][3]} \cdot A^8
	\\ \\
	0&0&0&0&0 & \left( [11]+\frac{[20]}{[4]}  \right) \cdot A^8
	
\end{array}\right).
	\nn
	\ee
}

\noindent
This triangular transformation is immediately converted into a formula for
${\cal F}$:
\be
H_R^{{\rm family}_a(k)} = \sum_{X\subset R}
{\cal Z}^{X\,{\rm family}_a}_{R}\cdot { F}_X^{{\rm twist}_k}
= \sum_{X\subset R}
{Z}^{X}_{R}\cdot {\cal F}_X^{\,{\rm family}_a(k)}.
\label{ZcalZ}
\ee
Universality of the matrix $V$ implies universality of the differential expansion ---
the fact that its coefficients ${\cal F}_X$ are independent of $R$.

We obtained this {\bf new} triangular property in the example of ``family$_a$'',
but it can be true in a much wider context -- perhaps, even beyond the
two-bridge arborescent knots.
And this can be {\bf the simplest way to justify the existence of
the universal differential expansion}.

\bigskip

According to (\ref{ZcalZ}) the properties ${ F}_X^{{\rm twist}_{-1}}=1$,
${ F}_X^{{\rm twist}_0}=\delta_{X,\emptyset}$,
${ F}_X^{{\rm twist}_1}= \Lambda_X^{\rm trefoil}$
of the twist functions
${ F}_X^{{\rm twist}_{k}} = \left<X|{\cal B}^{k+1}U|\emptyset\right>
= \sum_Y \left<X|{\cal B}^{k+1}|Y\right>$
imply that
\be
\sum_{X\subset R} {\cal Z}^{X\,{\rm family}_a}_{R} =
H_R^{{\rm family}_a(k=-1)} = H_{R}^{6_2}, \nn \\
{\cal Z}^{\emptyset\,{\rm family}_a}_{R} = H_R^{{\rm family}_a(k=0)} = H_{R}^{{\rm trefoil}}, \nn \\
\sum_{X\subset R} {\cal Z}^{X\,{\rm family}_a}_{R}\cdot \Lambda_X^{\rm trefoil} =
H_R^{{\rm family}_a(k=1)} = H_{R}^{5_1}.
\ee

Actually, the choice of matrix $V$ is ambiguous:  we can shift the content between the
last and penultimate columns, associated with representations $[2,1]$ and $[2,2]$.
Our choice was the maximal simplicity and absence of poles in the last column,
i.e. we assume the $F_{[2,1]}$ can have $D_1D_{-1}$ poles, but $F_{[2,2]}$ is just a polynomial. Such assumptions appear to provide reasonably nice elements of $V$.
Moreover, a simple calculation shows that $F_{X2}$ has a pole $[3]\{q\}^2$
(or that the $Z_{X2}$ for defect one is reduced to $[3]\{q\}^2D_{2}D_{-2}$).
This is what we postulate/conjecture as the modification of DE for defect one,
and {\bf this is the second new result of this letter}.
It is intimately related to an unexpected new triangular structure --- the matrix $V$.
In our example it was related to the peculiarity of ``family$_a$'',
but one can now look for it far beyond.
{\bf This triangular structure is the third new result}.

\section{Non-rectangular $R=[r_1,r_2]$, the first two levels \label{general}}

The current situation with non-rectangular representations is reviewed in \cite{M20},
where numerous references are given for development of the story, with a long history
of insights and errors.
In this section we make a new attempt to find the generalization of (\ref{recZ})
for the $Z$-factors to non-rectangular $R$
and describe a new puzzle, which needs to be resolved.

\subsection{The case of $R=[r,1]$}

Today we seem to know the structure of DE for all $R=[r,1]$,
and in the above text we demonstrated that it works for generic knots, even with defects.\footnote{
Everywhere in the present letter this means not a proof or even generic explanation,
but just a demonstration of how ``impossible'' things still happen to be true in particular examples.}
For $R=[r,1]$  the summation domain $M_R$ consists of a set of the composite representations
\be
M_{[r,1]} = [r,1]\otimes\overline{[r,1]}
= id +  \sum_{i=1}^{r-1} \Big(2 ([i],[i]) + ([i,1],[i+1])+([i+1],[i,1])\Big)
+ ([r],[r]) +\sum_{i=1}^r ([i,1],[i,1])
\label{Mr1}
\ee
Of main interest is the first big sum, which contains symmetric composites $([i],[i])$
twice (with multiplicity two) and involves also non-diagonal composites,
which are pairwise equal and will be denoted by $X_{i+1}:=([i,1],[i+1])+([i+1],[i,1])$
for brevity.
For differential expansion this means that
\be
H_{[r,1]}^{\cal K} = 1 +  \sum_{i=1}^{r-1}
\left(\Big(Z_{[r,1]}^{[i]'} + Z_{[r,1]}^{[i]''}\Big)\cdot F_{[i]}^{\cal K}
+ Z_{[r,1]}^{X_{i+1}}\cdot F_{X_{i+1}}\right) + Z_{[r,1]}^{[r]}\cdot F_{[r]}^{\cal K}
+ \sum_{i=1}^r Z_{[r,1]}^{[i,1]}\cdot F_{[i,1]}^{\cal K}
\label{DEr1}
\ee
In \cite{M20} a nice interpretation of the $Z$-factors was given in terms of the
pentad matrix $U$.
We provide further details in the next subsection \ref{Uapp21}.
However, in subsection \ref{DEr1r2problems}
we will see that actual separation $Z'+Z''$ into factorized
items turns out to be different.
This change does not affect the DE (\ref{DEr1}) itself,
but makes its explanation questionable again.
We will see the problem by straightforward attempt to work out the $Z$-factors
for a bigger set of representations $R=[r_1,r_2]$.

\subsection{$U$-matrix approach for $R=[2,1]$ in detail
\label{Uapp21}}

As shown in \cite{M20}, the $10\times 10$ matrices in the case of $R=[2,1]$ have a block form, and the $2\times 2$ block can be basically ignored in consideration
of Racah matrices (it is trivially restored).
Moreover, in twist knot calculus the matrices can be even reduced to $6\times 6$
or at least $7\times 7$ (if one wants to have all the $Z$-factors nicely factorized).
However, in this letter we do not go so far in reductions, and consider the
truly interesting $8\times 8$ matrices.
The universal triangular pair is:

{\footnotesize
\be
\!\!\!\!\!\!\!\!\!\!\!\!\!\!\!\!\!\!\!
B_{[2,1]}=\left(\begin{array}{cccccccc}
1 \\ \\
0 & A^2 \\ \\
-A^2 & 0 & A^2 \\ \\
-A^2 & 0 & 0 & A^2 \\ \\
\frac{A^4}{q^2} & \frac{A^4}{ q^3D_0}\sqrt{\frac{D_2}{D_{-2}}}
&-\frac{A^4}{[2]q^3}& -\frac{[3]A^4}{[2]q^3} & \frac{A^4}{q^4} \\ \\
q^2A^4  & -\frac{q^3A^4}{D_0}\sqrt{\frac{D_{-2}}{D_2}}
&-\frac{q^3A^4}{[2]} & -\frac{[3]q^3A^4}{[2] } &0& q^4A^4  \\ \\
-A^6& -\frac{[3]A^5\{q\}}{ D_0\sqrt{D_2D_{-2}}}&\frac{[3]A^6}{[2]^2} & \frac{[3]^2A^6}{[2]^2}
& -\frac{[3]A^6}{q[2]}& -\frac{[3]qA^6}{[2]} & A^6 \\ \\
-A^6 & -\frac{[3]A^5\{q\}}{ D_0\sqrt{D_2D_{-2}}}+\frac{A^6}{\{q\} \sqrt{D_2D_{-2}}}
& \frac{[3]A^6}{[2]^2}-\frac{A^5D_0}{[2]^2\{q\}^2}
& \frac{[3]^2A^6}{[2]^2} + \frac{A^5D_0}{[2]^2\{q\}^2}
& -\frac{[3]A^6}{q[2]}+\frac{q^2A^5D_0}{[2]\{q\}}& -\frac{[3]qA^6}{[2]}-\frac{A^5D_0}{q^2[2]\{q\}}
& 0 & A^4
\end{array}\right)
\nn
\ee
}
\noindent
and

{\footnotesize
	
		\be
	\frac{{\cal E}_{[2,1]}}{d_{[2,1]}}=
	\nn
	\ee
}

{\tiny
	\be
	\!\!\!\!\!\!\!\!\!\!\!\!\!\!\!\!\!\!\!\!\!\!\!\!\!\!\!\!\!\!\!\!\!\!\!
	\left(\begin{array}{cccccccc}
		1 \\ \\
		0 & \frac{\{q\}}{A\sqrt{2D_1D_{-1}}} \\ \\
		\frac{A}{D_0} & 0 & -\frac{\{q\}}{AD_0\sqrt{D_1D_{-1}}} \\ \\
		\frac{A}{D_0} & \frac{[2]^2\{q\}^2(A+A^{-1})}{[3]A\sqrt{2D_0 \prod_{i=-2}^2D_i}}
		& \frac{\{q\}}{[3]AD_0\sqrt{D_1D_{-1}}} & -\frac{[2]^2\{q\}}{[3]AD_0\sqrt{D_1D_{-1}}} \\ \\
		\frac{A^2}{qD_0D_{-1}} & -\frac{\{q\}}{q\sqrt{2D_0 \prod_{i=-2}^2D_i}} & 0
		& -\frac{[2]\{q\}\sqrt{D_1D_{-1}}}{q\prod_{i=-2}^1D_i}
		& \frac{q^3[2]\{q\}^2\sqrt{D_1D_{-3}}}{A^2\prod_{i=-3}^1D_i}\\ \\
		\frac{qA^2}{D_1D_0} & \frac{q\{q\}}{\sqrt{2D_0\prod_{i=-2}^2D_i}} & 0
		& -\frac{q[2]\{q\}\sqrt{D_1D_{-1}}}{\prod_{i=-1}^2D_i} & 0 & \frac{[2]\{q\}^2\sqrt{D_3D_{-1}}}{q^3A^2\prod_{i=-1}^3D_i} \\ \\
		\frac{A^3}{D_1D_0D_{-1}} & 0 & 0 & -\frac{[3]\{q\}A\sqrt{D_1D_{-1}}}{\prod_{i=-2}^2D_i}
		& \frac{q^4[3]\{q\}^2\sqrt{D_1D_{-3}}}{A\prod_{i=-3}^2D_i}
		& \frac{[3]\{q\}^2\sqrt{D_3D_{-1}}}{q^4A\prod_{i=-2}^3D_i}
		& -\frac{[3]\{q\}^3\sqrt{D_3D_{-3}}}{A^3\prod_{i=-3}^3D_i} \\ \\
		\frac{A^3}{D_1D_0D_{-1}} & -\frac{A^2+1}{[3]\sqrt{2D_0\prod_{i=-2}^2D_i}}
		& -\frac{A}{[3]\{q\}D_0\sqrt{D_1D_{-1}}}
		& -\frac{[4]\{q\}A\sqrt{D_1D_{-1}}}{[2]\prod_{i=-2}^2D_i}+
		&-\frac{q^4\sqrt{D_1D_{-3}}}{[2]A\prod_{i=-2}^1D_i}
		& -\frac{\sqrt{D_3D_{-1}}}{[2]q^4A\prod_{i=-1}^2D_i} & 0
		& \frac{2}{[2]A\sqrt{2D_0\prod_{i=-2}^2D_i}} \\
		&&&+\frac{A}{[3]\{q\}D_0\sqrt{D_1D_{-1}}}
	\end{array}\right),
    \nn
	\ee}
where $d_{[2,1]} =  \frac{D_1D_0D_{-1}}{[3]\{q\}^3}$.
\vfill

\noindent
$\mathcal{E}_{[2,1]}$ is the eigenvector matrix of $B_{[2,1]}$
\be
B_{[2,1]}{\cal E}_{[2,1]} = {\cal E}_{[2,1]}\bar T_{[2,1]}^2
\ \ \ \Longleftrightarrow \ \ \
{\cal E}_{[2,1]}^{-1}B_{[2,1]}{\cal E}_{[2,1]} = \bar T_{[2,1]}^2.
\ee
The next two matrices $\bar{S}_{[2,1]}$ and $U_{[2,1]}$ are no longer universal. Moreover, they
depend on the choice of the $Z$-factors (\ref{ZZ21}) and the normalization
of ${\cal E}_{[2,1]} \longrightarrow {\cal E}_{[2,1]}\cdot{\rm diag}(K_{[2,1]})$:
\be
\bar S_{[2,1]} = \bar T_{[2,1]}^2 {\cal E}_{[2,1]}^{\rm transp }\cdot
{\rm diag}\left(\frac{Z_{[2,1]}^Q}{\Lambda'_Q}\right)\cdot
 {\cal E}_{[2,1]}\bar T_{[2,1]}^2
 \label{SviaE21}
\ee
\be
U_{[2,1]} = {\cal E}_{[2,1]}\bar S_{[2,1]}^{-1}
\ee
The choice in \cite{M20} was directly related to (\ref{Z21})
\be
Z_{[2,1]}^\emptyset = 1, \ \  Z_{[2,1]}^{[1]} = [3],
\ \  Z_{[2,1]}^{[1]'} = \frac{[3]}{[2]^2}D_0^2,
\ \  Z_{[2,1]}^{[1]''} = \frac{[3]^2}{[2]^2}D_2D_{-2},
\ \ Z_{[2,1]}^{[1,1]} = \frac{[3]}{[2]}D_2D_0D_{-2}D_{-3},
\nn \\
  Z_{[2,1]}^{[2]} = \frac{[3]}{[2]}D_3D_2D_0D_{-2},
\ \  Z_{[2,1]}^{[2,1]} = D_3D_2D_1D_{-1}D_{-2}D_{-3},
\ \  Z_{[2,1]}^{[X2]} = -[3]^2\{q\}^4D_2D_{-2}
\label{ZZ21}
\ee
and
\be
\Lambda'_\emptyset = 1, \ \ \ \Lambda'_{[1]} = -A^2, \ \ \ \
\Lambda'_{[1,1]} = q^{-2}A^4, \ \ \ \Lambda'_{[2]} = q^2A^4, \ \ \ \
\Lambda'_{[2,1]} = \Lambda'_{[X2]} = -A^6.
\ee

Seven $Z$-factors contributing to DE of representation $R=[2,1]$ are then reproduced as
\be
Z_{[2,1]}^X = d_{[2,1]}\cdot\left<\emptyset\Big|\bar S_{[2,1]}\bar T_{[2,1]}^2 \bar S_{[2,1]}
\bar T_{[2,1]}^{-2} \bar S_{[2,1]} U_{[2,1]}^{-1}\Big| X \right>.
\label{matelZ21}
\ee
Note that $Z_{[2,1]}^{[1]}$ does not contribute to the DE of $[2,1]$,
because the corresponding matrix element of $U_{[2,1]}$ equels to zero:
$U_{[1],\emptyset} = 0$, while the others matrix elements, corresponding to representation $[1]$ are unities: $U_{[1]',\emptyset}=U_{[1]'',\emptyset}=1$. Together $Z_{[2,1]}^{[1]'}$ and $Z_{[2,1]}^{[1]''}$ reproduse (\ref{Z1r1r1}).
This fact looks non-trivial and we emphasize that among the elements (\ref{ZZ21}) used for  the calculation of $\bar{S}_{[2,1]}$ (\ref{SviaE21}) element  $Z_{[2,1]}^{[1]}$ does not vanish.

The $\bar S_{[2,1]}$ matrix, calculated from (\ref{SviaE21}), coincides with the
$8\times 8$ block of the one, found from direct Racah calculus in \cite{GuJo}
and later used in development of arborescent calculus and its extensions in
\cite{arbor,NRZ,3strfamilies}:
	\be
d_{[2,1]}	\bar S_{[2,1]} =
	\nn
	\ee
	{\tiny
\be
	\!\!\!\!\!\!\!\!\!\!\!\!\!\!\!\!\!\!\!\!\!\!\!\!\!\!\!\!\!\!\!\!\!\!\!\!\!\!\!
 \left(\begin{array}{cccccccc}
1 & 0 & \frac{\sqrt{D_1D_{-1}}}{\{q\}} & \frac{\sqrt{D_1D_{-1}}}{\{q\}} & \frac{\sqrt{D_1D_0^2D_{-3}}}{[2]\{q\}^2} & \frac{\sqrt{D_3D_0^2D_{-1}}}{[2]\{q\}^2} & \frac{D_1D_{-1}\sqrt{D_3D_{-3}}}{[3]\{q\}^3} & \frac{\sqrt{2D_2D_1D_{-1}D_{-2}}}{[2]\{q\}^2} \\ \\
\vdots & -\frac{2D_1D_{-1}}{[3]\{q\}^2} &0 & -\frac{(A^2+1)D_1D_{-1}}{[3]\{q\}A} \sqrt{\frac{2}{D_2D_{-2}}}& -\frac{D_1D_{-1}}{[3][2]\{q\}^3}\sqrt{\frac{2D_2D_{-1}D_{-3}}{D_{-2}}} &\frac{D_1D_{-1}}{[3][2]\{q\}^3}\sqrt{\frac{2D_3D_{1}D_{-2}}{D_{2}}} &0 & -\frac{(A^2+1)D_1D_{-1}}{[3][2]\{q\}^2A} \\ \\
&\vdots &- \frac{D_1D_{-1}}{[3]\{q\}^2}& \frac{[4]D_1D_{-1}}{[3][2]\{q\}^2} &\frac{D_2D_1\sqrt{D_{-1}D_{-3}}}{[3][2]\{q\}^3}& \frac{D_{-1}D_{-2}\sqrt{D_1D_3}}{[3][2]\{q\}^3} &0 & -\frac{D_1D_{-1}\sqrt{2D_2D_{-2}}}{[3][2]\{q\}^3} \\ \\
 &  &\vdots & \frac{D_1D_{-1}P_1}{[3]\{q\}^2D_2D_{-2}} & \frac{D_1\sqrt{D_{-1}D_{-3}}P_2}{[3][2]\{q\}^3D_2D_{-2}} & \frac{D_{-1}\sqrt{D_3D_1}P_3}{[3][2]\{q\}^3 D_2D_{-2}} &-\frac{D_1D_{-1}\sqrt{D_3D_1D_{-1}D_{-3}}}{\{q\}^2D_2D_{-2}} & \frac{D_1D_{-1}P_4}{[3][2]\{q\}^3}\sqrt{\frac{2}{D_2D_{-2}}} \\ \\
 & & &\vdots &-\frac{D_1D_0 P_5}{[3][2]^2\{q\}^2 D_2 D_{-2}} & \frac{[3]D_0^2\sqrt{D_3D_1D_{-1}D_{-3}}}{[2]^2\{q\}^2D_2D_{-2}} & \frac{D_1D_{-1}\sqrt{D_3D_1}}{\{q\} D_2D_{-2}} & -\frac{D_1}{[2]^2\{q\}^2}\sqrt{\frac{2D_2D_{-1}D_{-3}}{D_{-2}}}\\ \\
 & & & &\vdots &- \frac{D_0D_{-1}P_6}{[3][2]^2\{q\}^2D_2D_{-2}} & \frac{D_1D_{-1}\sqrt{D_{-1}D_{-3}}}{\{q\}D_2D_{-2}}  & -\frac{D_{-1}}{[2]^2\{q\}^2}\sqrt{\frac{2D_3D_1D_{-2}}{D_2}} \\ \\
 & & & & &\vdots &- \frac{D_1D_{-1}}{D_2D_{-2}} & 0\\ \\
 & & & & & &\vdots & \frac{2 D_1D_{-1}}{[2]^2\{q\}^2}
\end{array}\right),
\nn
\ee}

\noindent
where matrix $S_{[2,1]}$ is symmetric and the other half is easily restored. We also used the following notations to shorten the expression:
\be
\begin{array}{lll}
	P_1 =2A^2+\frac{2}{A^2}-\frac{[6][3]}{[2]}+5, &
	P_2 = A^3-\frac{1}{A^3} -[3]^2 \{q\}^2 D_0-A \frac{2q^6+1}{q^4} + \frac{q^6+2}{q^2 A},&
	P_3 = \left. P_2 \right|_{q\rightarrow q^{-1}}, \\
	P_4 = A^2+\frac{1}{A^2}-\frac{[8]}{[2]}+2, &
	P_5  = q A^2 + \frac{1}{q A^2} +3[2]-[8],&
	P_6 = \left. P_5\right|_{q\rightarrow q^{-1}}. \\
\end{array}
\ee

The $U_{[2,1]}$ is also complicated:

	\be
	U_{[2,1]} =
	\ee
	{\tiny
\be
	\!\!\!\!\!\!\!\!\!\!\!\!\!\!\!\!\!\!\!\!\!\!\!\!\!\!\!\!\!\!\!\!\!\!\!\!\!
 \left(\begin{array}{cccccccc}
1 & 0 & \frac{\sqrt{D_1D_{-1}}}{\{q\}} & \frac{\sqrt{D_1D_{-1}}}{\{q\}} & \frac{D_0\sqrt{D_1D_{-3}}}{[2]\{q\}^2} & \frac{D_0\sqrt{D_3D_{-1}}}{[2]\{q\}^2}& \frac{D_1D_{-1}\sqrt{D_3D_{-3}}}{[3]\{q\}^3} & \frac{\sqrt{2D_2D_1D_{-1}D_{-2}}}{[2]\{q\}^2} \\ \\
0&-\frac{ \sqrt{2D_1D_{-1}}}{[3]\{q\}A} &0 &-\frac{Q_{2,4}}{[3]}\sqrt{\frac{D_1D_{-1}}{D_2D_{-2}}} & -\frac{ D_{-1}\sqrt{D_2D_1D_{-3}}}{[3][2]\{q\}^2A\sqrt{D_{-2}}} & \frac{ D_1\sqrt{D_3D_{-1}D_{-2}}}{[3][2]\{q\}^2 A\sqrt{D_2}}  & 0& -\frac{Q_{2,4}\sqrt{D_1D_{-1}}}{\sqrt{2}[3][2]\{q\}} \\ \\
1& 0& \frac{Q_{3,3}\sqrt{D_1D_{-1}}}{[3]\{q\}D_0}& \frac{Q_{3,4}\sqrt{D_1D_{-1}}}{[3][2]\{q\}D_0} & \frac{Q_{3,5}\sqrt{D_1D_{-3}}}{[3][2]\{q\}^2 D_0} &\frac{Q_{3,6}\sqrt{D_3D_{-1}}}{[3][2]\{q\}^2 D_0}&\frac{A D_1D_{-1}\sqrt{D_3D_{-3}}}{[3]\{q\}^3D_0} & \frac{Q_{3,3}\sqrt{2D_2D_1D_{-1}D_{-2}}}{[3][2]\{q\}^2 D_0} \\ \\
1 & 0  & \frac{\sqrt{D_{1} D_{-1}}}{\{q\} }& \frac{Q_{4,4}\sqrt{D_{1} D_{-1}}}{  [3] \{q \} D_{2} D_{-2}} & \frac{Q_{4,5}D_ 0 \sqrt{D_{1} D_{-3}}}{[3] [2]\{q \}^2 D_ 2 D_{-2}    } &\frac{Q_{4,6}D_0 \sqrt{D_{3} D_{-1}}}{[3][2]\{q\}^2 D_{2} D_{-2}} &\frac{Q_{4,7}D_{1} D_{-1} \sqrt{D_{3} D_{-3}}}{[3]\{q\}^3D_{2} D_{-2} }& \frac{Q_{4,8}}{[3]^2[2]\{q\}^2} \sqrt{\frac{D_1D_{-1}}{2D_2D_{-2}}}\\ \\
1 & \frac{q\sqrt{2D_2D_1D_{-1}}}{[3]\{q\}A D_0\sqrt{D_{-2}}} & \frac{Q_{5,3}\sqrt{D_1D_{-1}}}{[3]\{q\}D_0} & \frac{Q_{5,4}\sqrt{D_1D_{-1}}}{[3]\{q\}D_2D_0D_{-2}} &  \frac{Q_{5,5}\sqrt{D_1} }{[3][2]\{q\}^2D_2D_{-2} \sqrt{D_{-3}} }  & \frac{Q_{5,6}D_0\sqrt{D_3D_{-1}}}{[2]\{q\}^2D_2D_{-2}} & \frac{Q_{5,7}D_1D_{-1}\sqrt{D_3}}{[3]\{q\}^3D_2D_0D_{-2}\sqrt{D_{-3}}} & \frac{Q_{5,8}}{[3][2]\{q\}^2D_0}\sqrt{\frac{D_1D_{-1}}{2D_2D_{-2}}} \\ \\
1 & -\frac{\sqrt{2D_1D_{-1}D_{-2}}}{[3]q\{q\}AD_0\sqrt{D_2}} &\frac{Q_{6,3}\sqrt{D_1D_{-1}}}{[3]\{q\}D_0} & \frac{Q_{6,4}\sqrt{D_1D_{-1}}}{[3]\{q\}D_2D_0D_{-2}}& \frac{Q_{6,5}D_0\sqrt{D_1D_{-3}}}{[2]\{q\}^2D_2D_{-2}}& \frac{Q_{6,6}\sqrt{D_{-1}}}{[3][2]\{q\}^2D_2D_{-2}\sqrt{D_3}}& -\frac{Q_{6,7}D_1D_{-1}\sqrt{D_{-3}}}{[3]\{q\}^3D_2D_0D_{-2}\sqrt{D_3}} &\frac{Q_{6,8}}{[3][2]\{q\}^2D_0} \sqrt{\frac{D_1D_{-1}}{2D_2D_{-2}}}\\ \\
1 &\sqrt{\frac{2D_1D_{-1}}{D_2D_0^2D_{-2}}} &\frac{A\sqrt{D_1D_{-1}}}{\{q\}D_0} & \frac{Q_{7,4}\sqrt{D_1D_{-1}}}{\{q\}D_2D_0D_{-2}}& \frac{Q_{7,5}\sqrt{D_1}}{[2]\{q\}^2D_2D_{-2}\sqrt{D_{-3}}} &-\frac{Q_{7,6}\sqrt{D_{-1}}}{[2]\{q\}^2D_2D_{-2}\sqrt{D_3}} &\frac{Q_{7,7}}{[3]\{q\}^3D_2D_0D_{-2}\sqrt{D_3D_{-3}}} & \frac{Q_{7,8}}{[2]\{q\}^2D_0}\sqrt{\frac{2D_1D_{-1}}{D_2D_{-2}}} \\ \\
1 & \frac{Q_{8,2}}{[3]\{q\}^2}\sqrt{\frac{2D_1D_{-1}}{D_2D_0^2D_{-2}}}& \frac{Q_{8,3}\sqrt{D_1D_{-1}}}{[3]\{q\}^3D_0} & \frac{Q_{8,4}\sqrt{D_1D_{-1}}}{[3]\{q\}D_2D_0D_{-2}} & \frac{Q_{8,5}\sqrt{D_1D_{-3}}}{[3][2]\{q\}^2D_2D_{-2}} & \frac{Q_{8,6}\sqrt{D_3D_{-1}}}{[3][2]\{q\}^2D_{2}D_{-2}} & \frac{Q_{8,7}D_1D_{-1}\sqrt{D_3D_{-3}}}{[3]\{q\}^3D_2D_0D_{-2}} & \frac{Q_{8,8}}{[3]^2[2]\{q\}^4D_0}\sqrt{\frac{D_1D_{-1}}{2D_2D_{-2}}} \\
\end{array}\right),
\nn
\ee
}
 where
{\tiny
\begin{minipage}[t]{0.5\textwidth}
\be
\begin{array}{lll}
	Q_{2,4} &=& 1+A^{-2}, \\
	Q_{3,3} &=& [3]A+A^{-1}, \\
	Q_{3,4} &=& [3][2]A-[4]A^{-1}, \\
	Q_{3,5} &=& [3]A D_0 -A^{-1}D_{2}, \\
	Q_{3,6} &=& [3]A D_0 -A{-1} D_{-2} ,\\
	Q_{4,4} &=& [3]A^2+1-[8][3][4]^{-1} + [4][2]^{-1}A^{-2}, \\
	Q_{4,5} &=& [3]A^2+1-[3]q^4-[2]q^{-5}+q^{-2}A^{-2},\\
	Q_{4,6} &=& [3]A^2+1-[3]q^{-4}-[2]q^5+q^2A^{-2}, \\
	Q_{4,7}&=&A^2-[2][4]^{-1}[6]^{-1}[12],\\
	Q_{4,8}&=&  2[3]^2A^2 -5[8][4]^{-1}-2[2]q^{-7}(1+q^{6}+q^{{10}}+q^{14})- \\  &&2q^{-6}(1-q^{10}+q^{12})+([4][2]+2)A^{-2}, \\
		Q_{5,3} &=&[3]A-q^{-2}A^{-1}, \\
	Q_{5,4}&=& [3]A^3-q^{-6}(1+q^2+2q^4-q^6+q^{10}+q^{12})D_0-\\&&(1+2q^2-q^4)A-A^{-3}, \\
	Q_{5,5} &=& [3]q^{-3}A^4-q^{-9}A (q^{14}+q^{12}+q^{10}+q^8-q^6+ \\&&q^4+q^2+1)D_0  -q^3 A^2+q^{-7}(q^{16}-2 q^{10}+q^8-1)-\\&&[6]([2][3]qA^2)^{-1},\\
		Q_{5,6}&=&q^{-4}(A^2 q^4-q^8+q^6-1), \\
	Q_{5,7}&=&q^{-3}A^4 +q^{-7}(A^2 (-q^{10}-q^8+q^6-1)+q^{14}-q^{12}+\\&&q^8-q^2+1),\\
	Q_{5,8}&=&2[3]A^3+q^{-6}(A^{-1}(2 q^8-q^6+q^4+2)-\\&&A(2 q^{12}+2 q^8-q^6+3 q^4+2 q^2+2)), \\
		Q_{6,3} &=& [3]A-q^2A^{-1}, \\
	Q_{6,4} &=& [3]A^3 -q^{-6}(A(q^{12}+q^{10}+2 q^8+2 q^4+1)-\\&& A^{-1}(q^{12}+q^{10}+2 q^8-q^6+q^2+1)),
	\\ \\
\end{array}
\nn
\ee
\end{minipage}
\begin{minipage}[t]{0.5\textwidth}
	\be
	\begin{array}{lll}
			Q_{6,5}&=&A^2 - q^{-4}(q^8-q^2+1),\\
		Q_{6,6}&=&[3]q^3A^4 - [6]q([3][2]A^2)^{-1} -q^{-5}A^2(q^{14}+q^{12}+\\&& q^{10}-q^8+q^6+q^4+2 q^2+1)+q^{-9}(q^{18}+q^{14}-\\&&q^{12}+q^{10}+2 q^8-q^6+q^4+1),\\
			Q_{6,7}&=&-q^{-7}(A^4 q^{10}-(A^2-1) q^{14}+A^2
		q^8-(A^2-1) q^6- \\&&A^2 q^4-q^{12}-q^2+1) ,\\
			Q_{6,8}&=&2[3]A^3-(q^6 A)^{-1}(A^2 (2 q^{12}+2 q^{10}+3 q^8-q^6+2
		q^4+2)\\&&-q^4 (2 q^8+q^4-q^2+2)),\\
			Q_{7,4} &=& A^3-[6][4]A([3][2]^2)^{-1}+[12][2]([6][4]A)^{-1}, \\
		Q_{7,5} &=& q^{-3}A^4 -q^{-7}A^2(q^{10}+q^8-q^6+q^4-q^2+1)+\\&&q^{-7}(q^{14}-q^{12}+q^{10}-q^8+2 q^6-q^4-q^2+1),\\
			Q_{7,6} &=& -q^3 A^4 +q^{-3}A^2(q^{10}-q^8+q^6-q^4+q^2+1)-\\&&q^{-7}(q^{14}-q^{12}-q^{10}+2 q^8-q^6+q^4-q^2+1),\\
		Q_{7,7} &=& A^7-[10][5]^{-1}[2]A^5+[12][2]^{-1}A^3 - q^{-12}A(q^{24}-q^{22}+\\&&q^{20}-q^{18}+4 q^{16}-3 q^{14}+2
		q^{12}-3 q^{10}+4 q^8-q^6+q^4-\\&&q^2+1) +[14][10]([7][5][2]^2A)^{-1}, \\
		Q_{7,8} &=& A^3-[10] ([5][2])^{-1}A, \\
		Q_{8,2}&=& -A^{-2}+[10]([5][2])^{-1}, \\
		Q_{8,3}&=& -A^{-1} [10]([5][2])^{-1} A,\\
			Q_{8,4}&=&[3]A^3 -[6][4][2]^{-2}A+[10][2]([5]A)^{-1} -A^{-3}, \\
		Q_{8,5} &=& [3]A^3 -q^{-6}A (q^{12}+q^8-q^6+2 q^4+1)+(q^6A)^{-1},\\
		Q_{8,6} &=& [3]A^3 - q^{-6}A (q^{12}+2 q^8-q^6+q^4+1) + q^6 A^{-1}, \\
			Q_{8,7} &=& A^3 - [10]([5][2])^{-1} A, \\
		Q_{8,8}  & = & q^{-10}A^{-1} (A^4 (2 q^{12}+q^8-2 q^6+q^4+2) q^4-\\&&2 A^2
		(q^{20}-q^{18}+2 q^{16}-3 q^{14}+4 q^{12}-2
		q^{10}+4 q^8-3 q^6+\\&&2 q^4-q^2+1)+(2
		q^{12}+q^8-2 q^6+q^4+2) q^4). \\	\\
	\end{array}
\nn
	\ee
\end{minipage}

}

We remind that both $\bar S_{[2,1]}$ and $U_{[2,1]}$ are not universal like
KNTZ matrix ${\cal B}_{[2,1]}$ and its eigenvector matrix ${\cal E}_{[2,1]}$,
i.e. they need to be calculated again for a new representation $R$.
Still, one can hope one day to get a general expression for these matrices,
comparable in ``simplicity'' to the hypergeometric formulas for (anti)symmetric representations $R$.

\subsection{Problems with generalization to $R=[r_1,r_2]$
\label{DEr1r2problems}}

We now proceed to generic two-line representations, and describe the new problems,
which arise in this case.
The usual way to find $Z$-factors is to use HOMFLY-PT polynomials of torus knots,
which can be calculated for any representation
from the Rosso-Jones formula \cite{RJ}, and then apply it to the particular case
of the trefoil ${\cal K}= 3_1$, where all
$F^{3_1}_Q =\Lambda^{\rm trefoil}_Q :=(-A^2)^{|Q|}q^{2\varkappa_Q}$
are explicitly known (the situation would be even simpler for ${\cal K}=4_1$ with
all $F^{4_1}_Q=1$, but there is no {\it a priori} explicit answer for
its colored polynomials, because its simplest representation is a three-strand braid (some polynomials can be found \cite{|R|=4} and \cite{knotebook}).
Then, knowing the l.h.s. of (\ref{de}) and having just a combination of $Z$-factors
on the r.h.s., we can try to find them from factorization condition.
We will now illustrate the first steps of this strategy.

The analogue of (\ref{Mr1}) for the generic two-column $R$ is more involved,
it is partly presented in \cite{M20}, but we actually checked it for a much larger
variety of representations.
Now we keep just the first terms:
\be
M_{[r_1,r_2]} = [r,1]\otimes\overline{[r,1]}
= id + 2([1],[1]) + 3([2],[2]) + ([1,1],[1,1]) + \ldots
\label{Mr1r2}
\ee
what implies DE in the form
\be
H_{[r_1,r_2]}^{\cal K} = 1 +  \Big(Z_{[r_1,r_2]}^{[1]'} +  Z_{[r_1,r_2]}^{[1]''}\Big)\cdot F_{[1]}^{\cal K}
+ \Big(Z_{[r_1,r_2]}^{[2]} +  Z_{[r_1,r_2]}^{[2]'}+  Z_{[r_1,r_2]}^{[2]''}\Big)\cdot F_{[2]}^{\cal K}
+ Z_{[r_1,r_2]}^{[1,1]}\cdot F_{[1,1]}^{\cal K} + \ldots
\label{der1r2}
\ee
with $F$-functions defined from the study of (anti)symmetric representations.
In fact, multiplicity $3$ drops down to $2$ for $r_2=1$ and $r_2=r_1-1$
and both multiplicities disappear in rectangular cases   $r_2=0$ and $r_2=r_1$.
This means that one of the three $Z^{[2]}$ factors should be proportional to $r_2-1$,
while another -- to $r_1-r_2-1$.
Likewise one of the $Z^{[1]}$ factors should be proportional to $r_2$, another to $r_2-r_1$.
We expect $Z_{Q}$ for all diagonal composites to be products of $2|Q|$ differentials,
i.e. to be of  the order $\{q\}^{2|Q|}$ when $A=q^k$ with any $k$.
$Z$-factors for non-diagonal composites contain extra factors of $\{q\}^4$
and are not expected to contribute up to the order $\{q\}^6$.

Explicit calculation for trefoil and, in fact, for arbitrary torus knot
(since we expect this to be true for {\it all} knots, not obligatory torus)
gives:
\be
\left.\frac{H_{[r_1,r_2]}^{{\cal K}} - 1}{ F_{[1]}^{{\cal K}}\cdot \{q\}^2 }\right|_{A=q^k}
 = (r_1-r_2)(k+r_1-r_2)(k-1) + 2r_2(k+r_1)(k-2) + O(\{q\}).
\ee
The splitting in two terms is consistent (or, perhaps, disctated) with vanishing conditions
at $r_2=0$ and $r_2=r_1$.
We can now get rid of $k$ and quantize this relation to get
\be
 Z_{[r_1,r_2]}^{[1]'} +  Z_{[r_1,r_2]}^{[1]''} =
[r_1-r_2]D_{r_1-r_2}D_{-1} +  \frac{[r_2][2(r_1-r_2+1)]}{[r_1-r_2+1]} D_{r_1}D_{-2} + O(\{q\}^4)
\label{Z1r1r1}
\ee
Note that after quantization the accuracy should increase to $O(\{q\}^4)$.
If we restrict ourselves to the particular case of  $r_2=1$, then we loose vanishing conditions,
and there is another option for splitting of the same quantity:
\be
[r-1]D_{r-1}D_{-1} + \frac{[2r]}{[r]}D_{r}D_{-2} = \frac{[r+1]}{[2][r]}\Big(
[r-1]D_{r-2}D_0   + [r+1] D_{r}D_{-2}\Big)
\label{Zr1}
\ee
The second one was deduced from the pentad study in \cite{M20},
but now we can suspect that correct is rather the first one --
though it does not admit a straightforward $U$-matrix formulation
(the freedom to play with is diagonal rescaling
${\cal E}\rightarrow {\cal E}\cdot{\rm diag}(K)$ however allows to adjust $\mathcal{E}$
to match the modified $Z$-factors, which was done for representation $R=[2,1]$ in the section  \ref{Uapp21}).
Note that even in (\ref{Z1r1r1})  the right quantization rule
$2\longrightarrow \frac{[2r]}{[r]}$ at its l.h.s.
is dictated by equality with the r.h.s. of (\ref{Zr1}).

Now we can proceed to the next order,  keeping in mind that the difference depends on the choice
of quantization in (\ref{Z1r1r1}). Two representations $[2]$ and $[1,1]$ contribute to the next order, and one can deduce the expression for $Z_{[r_1,r_2]}^{[1,1]}$, it consists of only one term:
\be
Z_{[r_1,r_2]}^{[1,1]} = \frac{[r_1+1][r_2]}{[2]}D_{r_1}D_{r_2-1}D_{-2}D_{-3}.
\ee

Splitting the rest between three terms, corresponding to representation $[2]$, turns out to be a  challenging problem which requires further investigation.

\section{Conclusion}

This letter describes the present situation and the newest achievements
in the subject of Differential Expansion of colored knot polynomials,
which is a quantum deformation of binomial expansion for special polynomials
at $q=1$, where
\be
H_R^{\cal K}(q=1,A)=\Big(H_{[1]}^{\cal K}(q=1,A)\Big)^{|R|}=
\Big(1+F_{[1]}^{\cal K}(q=1,A)\cdot \{A\}^2\Big)^{|R|}.
\ee
There is no {\it a priori} reason for such a deformation to $q\neq 1$ to exist beyond single-line
or single-row $R$, i.e. beyond (anti)symmetric coloring ---
nothing to say about exact expression.
Still, spectacular theory is already developed for twist knots.
We provided important evidence that those results can be extended to arbitrary knots.
Namely, the expansion remains just the same (structurally) for defect zero knots,
and we explained how to look for modifications in the case of knots with non-trivial defects:
first study an example of two-bridge family by just the same method, which was developed
for twist knots, observe a spectacular new triangle structure and then employ it
for extension beyond two bridges.
Finally we discussed the non-trivial $Z$-factors from exhaustive knowledge of trefoil,
which is not only a twist knot, but a torus knot and thus has all colored knot polynomials
immediately available.
Still extraction of DE structure for polynomials of trefoil  is a highly non-trivial task,
but it can be resolved in steps, and we make a new important step on this line.

One can now consider in the same way various knots with different defects, various
knot families and, most importantly, more complicated representations than the simplest
rectangular $[2,2]$ and non-rectangular $[2,1]$ in this letter.
Only full self-consistent picture with all representations $R$ involved will provide
the conclusive evidence for DE and justify the choices, which one needs to make in
particularly restricted cases.
Still, this letter illustrates once again that new steps can be made and keep DE
revived even when problems and doubts are mounting.
Once again doubts are resolved and the road is open towards new challenges.

\section*{Acknowledgements}

This work was partly supported by the Russian Science Foundation (Grant No.16-12-10344).

\end{document}